\documentclass[8.5pt,twoside,twocolumn]{article}
 \pdfoutput=1
\usepackage[super,sort&compress,comma]{natbib}
\usepackage[version=2]{mhchem}

\usepackage{times,mathptmx}
\usepackage{sectsty} 
\usepackage{balance} 
\usepackage{textcomp}

\usepackage{subfigure}
 \usepackage{graphicx} 
 \usepackage{lastpage}
\usepackage[format=plain,justification=raggedright,singlelinecheck=false,font=small,labelfont=bf,labelsep=space]{caption} 
\usepackage{fancyhdr}
\begin{document}

\twocolumn[
\begin{@twocolumnfalse} \noindent\LARGE{\textbf{Fluidic Integration of
Nanophotonic Devices Using Decomposable Polymers 
}} 
\vspace{0.6cm}
\noindent\large{\textbf{\\Ehsan Shah Hosseini,\textit{$^{a}$} Mehrsa
Raeis Zadeh,\textit{$^{b}$} Paul Kohl,\textit{$^{b}$} and Ali
Adibi$^{\ast}$\textit{$^{a}$}}}\vspace{0.5cm}

	%
	%
	\noindent\textit{\small{\textbf{Received Xth XXXXXXXXXX 20XX,
	Accepted Xth XXXXXXXXX 20XX\newline First published on the web Xth
	XXXXXXXXXX 200X}}}

	\noindent \textbf{\small{DOI: 10.1039/b000000x}} 
	%
	
	\noindent \normalsize{ Polynorbornene-based decomposable polymer which can be patterned  with ultraviolet  or electron-beam radiation
	is used to create micrometer-scale fluidic channels. Silicon nitride
	substrates are used to fabricate nanophotonic wavegide and
	resonators operating in the visible range of the spectrum. Fluidic
	channels generated by thermally decomposing the polymer through the
	oxide cladding is used to deliver ultra-small amounts of florescent
	samples to the optical sensors.  
	}
	\vspace{0.5cm} 
	\end{@twocolumnfalse}
]
\section{Introduction} 

%
\footnotetext{\textit{$^{a}$~ School of Electrical and Computer
Engineering, Georgia Institute of Technology, Photonics Research Group,
777 Atlantic Dr., Atlanta, GA 30332, U.S.A. Tel: +1 (404)385-3017;
E-mail: adibi@ece.gatech.edu
	}}
\footnotetext{\textit{$^{b}$~ School of Chemical and Biomolecular
Engineering, Georgia Institute of Technology, 311 Ferst Dr, Atlanta, GA
30332, U.S.A.
		}}
		%
		%
		%
		%
		%
		%
		%
		%
		%
		%
		%
	Micro/nanochannels have applications in various new technologies
	such as micro/nanofluidic devices.  Some key application areas for
	micro/nanofluidics are molecular biology, cellular biophysics, fuel
	cells, and photonics.  Specially, applications of micro/nanofluidics
	in biosensing are of interest.   Micro/nanofluidics technology
	allows for novel developments such as integration and multimode
	sensing.  Microfluidic circuitry can be mass-produced, making it
	inexpensive and accessible.  Moreover, the reduction of size greatly
	reduces the analysis time.  Another benefit of micro/nanosystems is
	the reduction in sample size needed.

	Incorporating advanced micro/nano fluidics with high-sensitivity
	photonic sensors will provide compact, effective sensors for
	lab-on-a-chip tools \cite{erickson2008nanobiosensors}.  Thus,
	optofluidic sensors are gaining widespread use in biosensing and
	chemical analysis applications \cite{nitkowski2002cavity}.  Some
	potential applications of optofluidic sensors are clinical
	screening, medical diagnostics, screening of chemical compounds in
	drug discovery and development, and toxic detection
	\cite{chao2006design}.

	The microfluidic integration of optical chips with the usually
	aqueous solutions can be done by three major methods: (1)  SU-8
	photolithography with glass/PDMS capping; (2) PDMS Replica Molding
	Process; (3) decomposable polymers. The first method uses SU-8 as
	the channel material. SU-8 is a high contrast, epoxy-based
	photoresist designed for micro-machining and other microelectronic
	applications, where a thick chemically and thermally stable image is
	desired.  The exposed and subsequently cross-linked portions of the
	film are rendered insoluble in liquid developers. SU-8 has very high
	optical transparency above 360 nm, which makes it ideally suited for
	imaging near vertical sidewalls in very thick films. SU-8 is best
	suited for permanent applications where it is imaged, cured and left
	in place. After the channels are defined using photolithography
	(with a dark-field mask) the liquid can be dropped on top of the
	reaction area or flown into the channels. If pressure driven flow
	(PDF) is required, channels can be covered either by a
	Polydimethylsiloxane (PDMS) layer or a glass cover. In the later
	case, access holes need to be etched through the glass cover.

	In the second method, a clear field mask is used to define the
	channel molds in SU-8.  SU-8 mold is made hydrophobic with a layer
	of Au evaporated and PDMS  is poured over the mold. After curing for
	2 hours at 80 \textdegree C  (during which the reservoirs can be
	incorporated in the film), the PDMS can be peeled off the mold. Then
	the PDMS piece should be made hydrophilic if a permanent and
	watertight structure is needed. To achieve this, the sample is
	exposed to an oxygen plasma in an RIE machine. It is shown
	\cite{bhattacharya2005studies}  that there is an optimum time for
	the exposure. If the RIE treatment is longer than 25 seconds the
	bond strength is degraded. If the sample remains exposed to air for
	a long time, a treatment in diluted (1:5) HCl is necessary before
	the oxygen plasma. After the oxygen treatment the PDMS surface
	retains its hydrophilic property for 15 minutes, which is enough for
	a proper alignment with the optical devices.

	\section{Theory} To flow the liquids into the channels there are two
	common methods. The first method (pressure driven flow or PDF)
	utilizes a pressure build up between the two reservoirs. The other
	major method is the electroosmotic flow (EOF).

	The generally required parameters of microfluidics, namely small
	size, small velocity and large viscosity, combine in devices to
	result in generally small values of an important dimensionless
	parameter, the Reynolds number \cite{tesar2007pressure}:
	\begin{equation} 
		\centering Re=w b / \nu, \end{equation} 
	where
	$w$[m/s] is the characteristic flow velocity, $b$[m] is the
	characteristic dimension and $\nu$[m\textsuperscript{2}/s] is
	kinematic viscosity  of the fluid. The $\nu$ parameter for water is
	1.01. $b$ is typically the smallest dimension along the channel. Due
	to the small dimensions of micro-channels, the $Re$ is usually much
	less than 100, often less than 1.0. In this Reynolds number regime,
	flow is completely laminar and no turbulence occurs. The transition
	to turbulent flow generally occurs in the range of Reynolds number
	2000. Laminar flow provides a means by which molecules can be
	transported in a relatively predictable manner through
	micro-channels. One of the basic laws of fluid mechanics for
	pressure driven laminar flow, the so-called no-slip boundary
	condition, states that the fluid velocity at the walls must be zero.
	This produces a parabolic velocity profile within the channel.
	Despite the simplicity of the pressure driven approach---which only
	needs a syringe pump or a vacuum line---the drawback is
	non-scalability of the devices. As for a rectangular channel with a
	characteristic dimension of $d$ and for a circular shaped tube with
	a radius $r$, the pressure needed for a certain velocity scales
	with: $$\Delta P \propto 1/wd^3$$ and $$\Delta P \propto 1/r^4.$$
	This imposes a limit on the size of the channels and makes
	nanofluicid with manageable pressures impossible. Therefore, if the
	channel sizes are smaller than roughly 10 $\mu$m the electroosmotic
	flow is the preferred method.

	EOF is the motion of liquid induced by an applied potential across a
	porous material, capillary tube, membrane, micro-channel, or any
	other fluid conduit. Because electroosmotic velocities are
	independent of conduit size, as long as the double layer is much
	smaller than the characteristic length scale of the channel,
	electroosmotic flow is most significant when in small channels.
	Therefore, in the smaller channels described in the following
	sections the flow is achievable  by applying a high voltage (200 V)
	through a pair of platinum electrodes across the LB conductive
	medium obtained from ``Faster Better Media LLC''.

\section{Fabrication}

Epoxy-functionalized polymers, such as polynorbornene (PNB), are
valuable for forming micrometer-size structures due to their case of
reaction.  Epoxy-based polymers can also be used as sacrificial
polymers.  PNB-based epoxy-containing decomposable polymers can be
exposed with ultraviolet (UV) or electron-beam (e-beam) radiation and
solvent developed to form free-standing structures.  In this work, a
negative-tone, PNB-based sacrificial material, identified as Unity
4698P, has been used to make microchannels.  Unity 4698P has a simple
process flow which can be accomplished in five process steps and can be
used to form arbitrarily, three dimensional (3D) shapes and channel
structures.  The developed sacrificial polymers patterns can be
encapsulated with a thick layer of silicon dioxide, which will not
affect the optical performance of the resonators
\cite{soltani2009novel}.  The thermal decomposition products of Unity
4698P are able to diffuse through the encapsulating silicon dioxide to
leave clean channels of an exact shape.

 \begin{figure}[htb] \centering \subfigure[]{
 \includegraphics[width=0.3\textwidth]{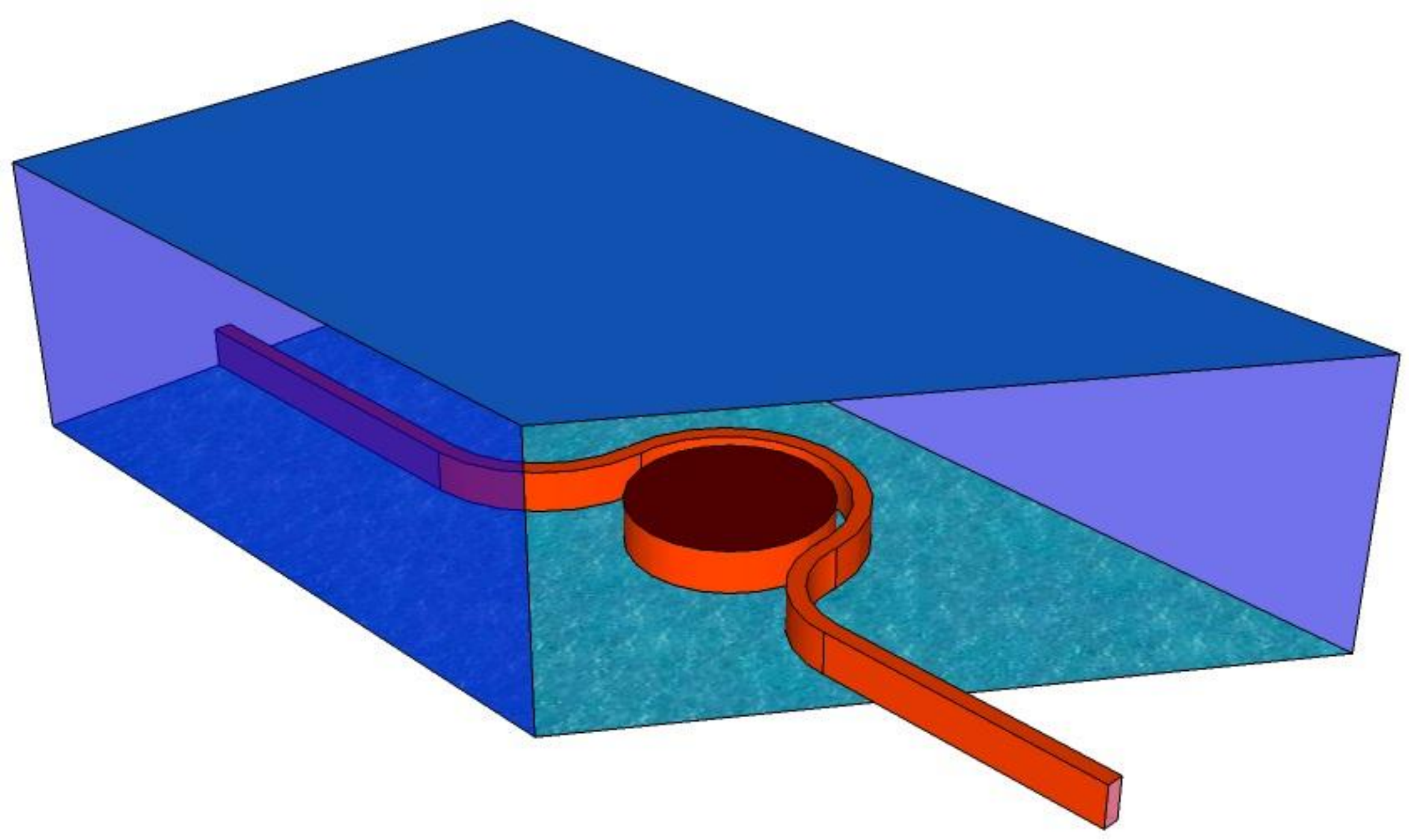} \label{fig:strct}
 }
 \subfigure[]{ \includegraphics[width=0.3\textwidth]{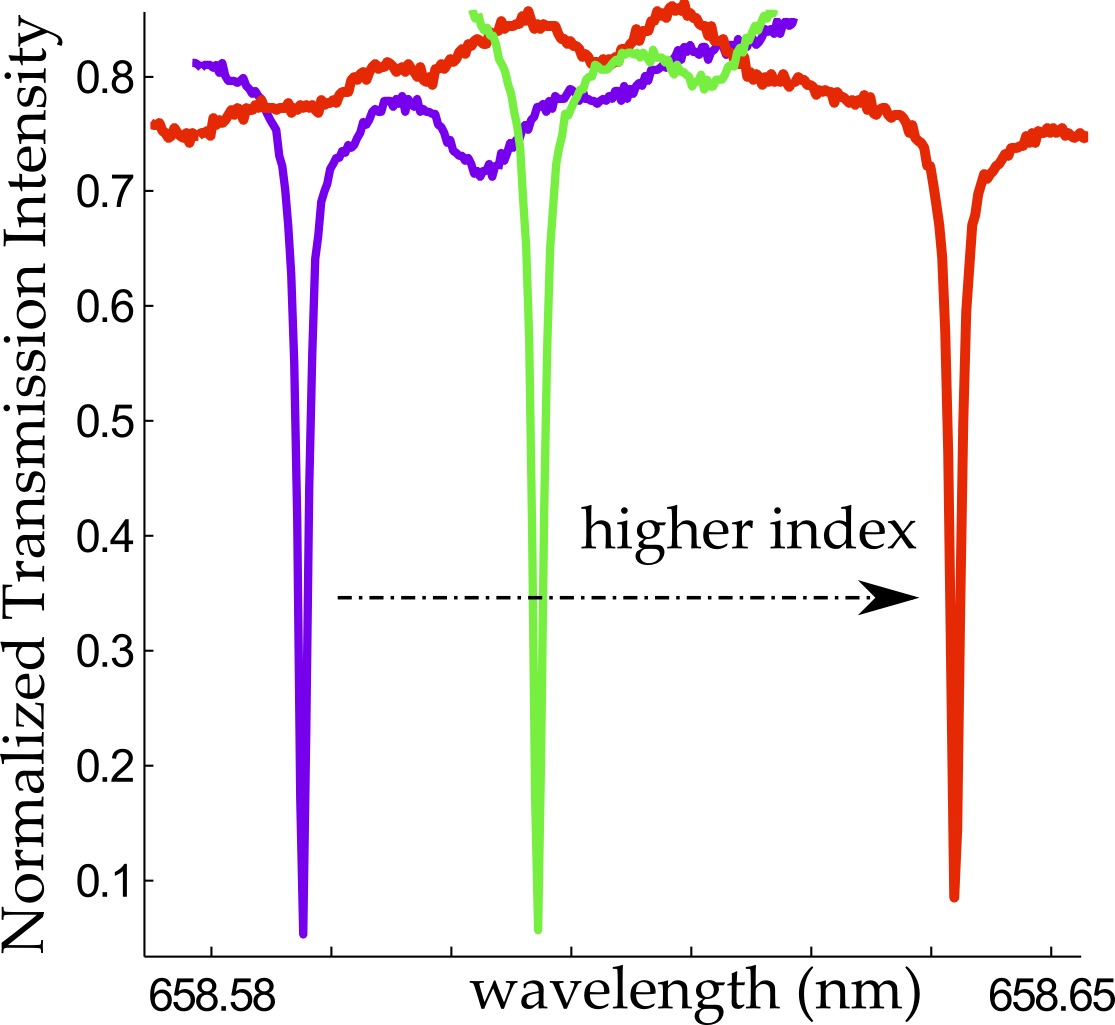}
 \label{fig:shift}
 }
  \label{fig:sensor} \caption[]{(a) Optofluidic sensor design with a
  straight waveguide coupled to a ring resonator and a microfluidic
  channel on top. The wavegide and resonator in this work are fabricated
  in silicon nitride and the channel is created by thermal decomposition
  of the polymer through the oxide cladding. (b) Change of the resonant
  frequency of the optical cavity due to the change in the refractive
  index of the ambient or selective attachment of the biomolecules to
  the resonator surface.} \end{figure}

  The design of the proposed optofluidic sensor is shown in Fig.
  \ref{fig:strct}.

   The device consists of a silicon nitride micro-ring resonator covered
   by a microfluidic channel.  The device is designed in a way that the
   evanescent light traveling in the ring resonator interacts with the
   upper fluidic cladding.  At resonance, light circulates many times
   within the ring, which leads to a large enhancement in the
   interaction length between the evanescent field and the cladding
   liquid (with micro/nanochannels, the interaction with the cladding
   fluid will be maximized \cite{soltani2009novel}.) There are several
   possible schemes for the sensing operation. In the refractive index
   sensing scheme, by injection of each refractive index fluid, the
   resonance spectrum of the microring resonator will vary.  Figure
   \ref{fig:shift} shows the variation of the resonance wavelength of
   the microring resonator when the refractive index of the fluid on top
   of the resonator is changed \cite{passaro2007ammonia}.

 High quality optical resonators can demonstrate very sharp resonances,
 which can be the core of very sensitive and accurate integrated sensing
 devices \cite{hosseini2009high}.

 \subsection{Fabrication of the photonic device} The first step of
 fabrication is to grow thick isolating oxides on prime silicon wafers.
 As the dry oxidation rate is very low, the majority of the oxidation
 process is done in a hydrogen rich environment (wet oxidation). The
 temperature during the oxidation is limited to 1100\textdegree C. The
 required time for the production of 3$\mu$m oxide is 32 hours. During
 this process 1.76$\mu$m of the silicon is consumed.

 The light guiding SiN layer can be deposited by either plasma enhanced
 chemical vapor deposition (PECVD) or low pressure chemical vapor
 deposition (LPCVD). The films deposited by LPCVD are of higher quality 
 than the PECVD films \cite{barclay2007fiber,yota2000comparative}. While
 PECVD SiN is etched approximately twice as fast by the inductively
 coupled plasma (ICP) reactive ion etching (RIE) dry etch as the LPCVD
 material, somewhat reducing the difficulty of the fabrication, the much
 higher impurity density \index{impurity density} (primarily hydrogen)
 and  higher optical absorption  makes it inappropriate for high quality
 devices \cite{yota2000comparative}. Nevertheless, the first microdisks
 fabricated were made in a PECVD chamber  demonstrated a moderate
 quality factor (Q$\sim10^5$).

 The photonic patterns are e-beam written with  MaN 2403 resist. MaN
 requires significantly lower dosage than the more common negative
 resist HSQ, therefore the writing time almost a quarter of that of the
 HSQ. A major issue with the MaN resist, nevertheless, is its weak
 adhesion to the substrate. The delamination failure, often happening in
 aqueous developers (e.g.  0.26N tetramethyl ammonium hydroxide
 developer (MF-319)  used in MaN's case), can be partially alleviated by
 hexamethyldisilazane (HMDS) priming. Nevertheless, the results with
 this method lack in repeatability. Instead, a  SurPass 3000 (provided
 by DisChem Inc.) is used \cite{cardenas2010wide} as a primer. The
 wafers are submerged in SurPass 3000 for 30 seconds, rinsed with
 deionized (DI) water and covered with MaN immediately. After the
 development of the exposed patterns an optional re-flow of the resist
 at 145\textdegree C for 3 minutes can lower the sidewall roughnesses
 significantly. This step, although very beneficial for ultra-high
 quality photonics, can lead to feature size changes or sticking of
 patterns if the MaN resist is not perfectly adhering to the substrate.

  After developing the patterns in MF-319 developer, the samples are
  etched in a \ce{CF4} ICP etcher with better than 1:1 selectivity
  \cite{hosseini2009high}. The resist is stripped away and samples are
  then cleaned in a piranha solution for 5 minutes. After dehydration
  samples are ready for fluidic integration described in \ref{fabfluid}.

  \subsection{Fabrication of the fluidic channels} \label{fabfluid}
  Unity 4698P is comprised of a polynorborene backbone with pendant
  epoxy moieties dissolved in 2-heptanone (Promerus LLC, Brecksville,
  OH).  The chemical structure of Unity 4698P is shown in Fig. 2.  When
  Unity 4698P is irradiated with UV radiation or an electron beam
  radiation, an acid catalyst is produced. Once reacted, the acid
  catalyst initiates epoxy ring opening and polymer cross-linking. An
  unstable carbocation from the epoxy which forms a covalent bond with
  other polynorbornene chains results in cross-linking between chains
  together. After the polymer has been cross-linked, Unity 4698P can be
  thermally decomposed at temperatures above 350\textdegree C.

    \begin{figure}[htb] \centering
    \includegraphics[width=0.4\textwidth]{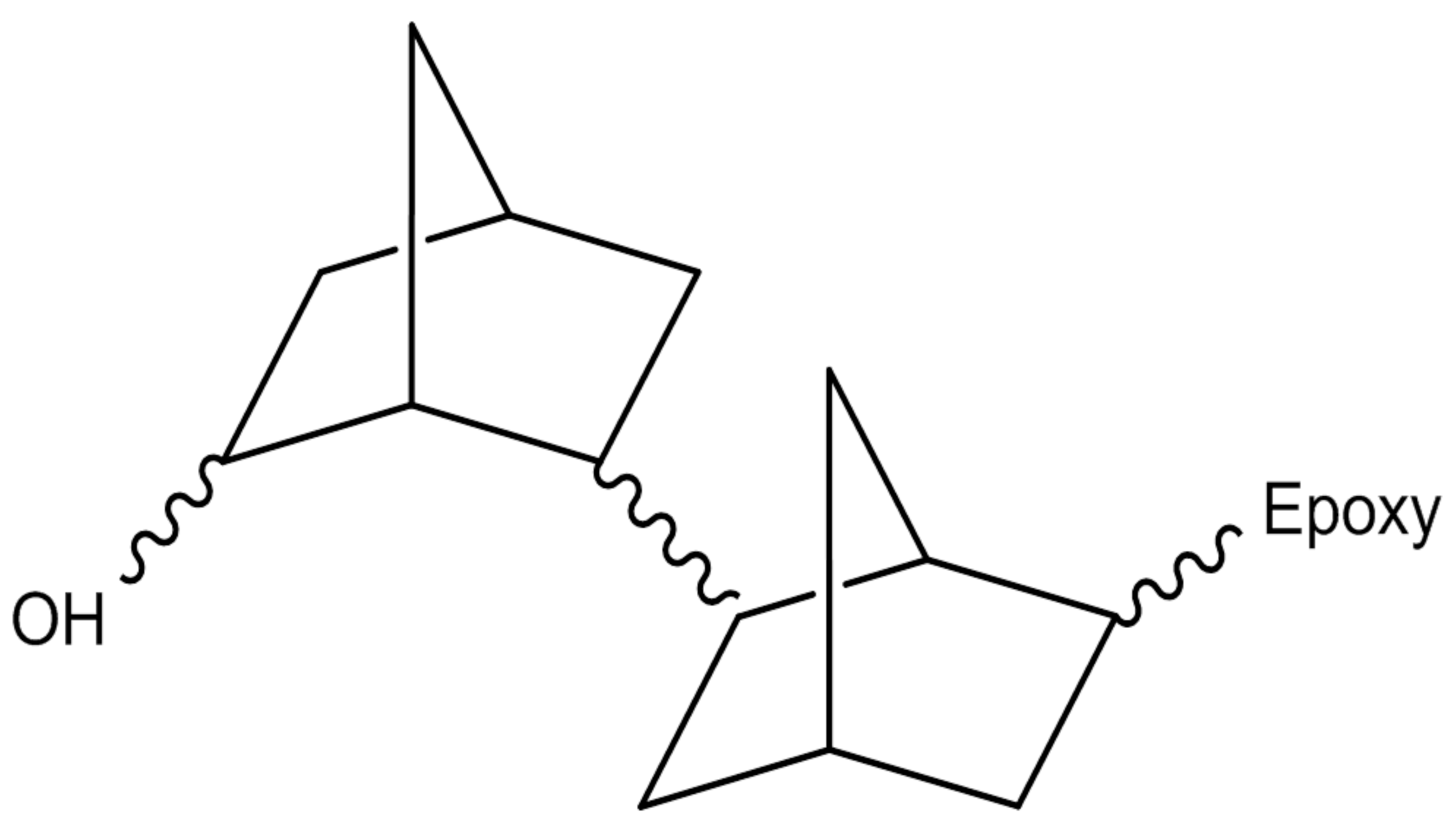}
    \caption[Polynorbornene]{The chemical structure of Unity 4698P.}
    \label{Polynorbornene} \end{figure}

  For microchannel fabrication, Unity 4698P films were spin-coated on
  the processed wafer described above using a Brewer Science CEE 100
  spinner.  A 8 to 9 $\mu$m thick film was obtained at a spin speed of
  500 rpm for 10 sec followed by 900 rpm for 60 sec. The films were
  soft-baked at 110\textdegree C for 5 min on a hot plate (air ambient)
  to remove most of the residual solvent.  The film thickness was
  measured after post-exposure bake using a Veeco Dektak profilometer.  
  In this work, e-beam lithography has been used to achieve very high
  spatial resolution. E-beam lithography was performed with a JEOL
  JBX-9300FS tool at 100 kV accelerating voltage and 50 pA beam current.
  After exposure, the samples were post-exposure baked at 90\textdegree
  C for 4 min on a hotplate.

\begin{figure}[htb] \centering
     \includegraphics[width=0.4\textwidth]{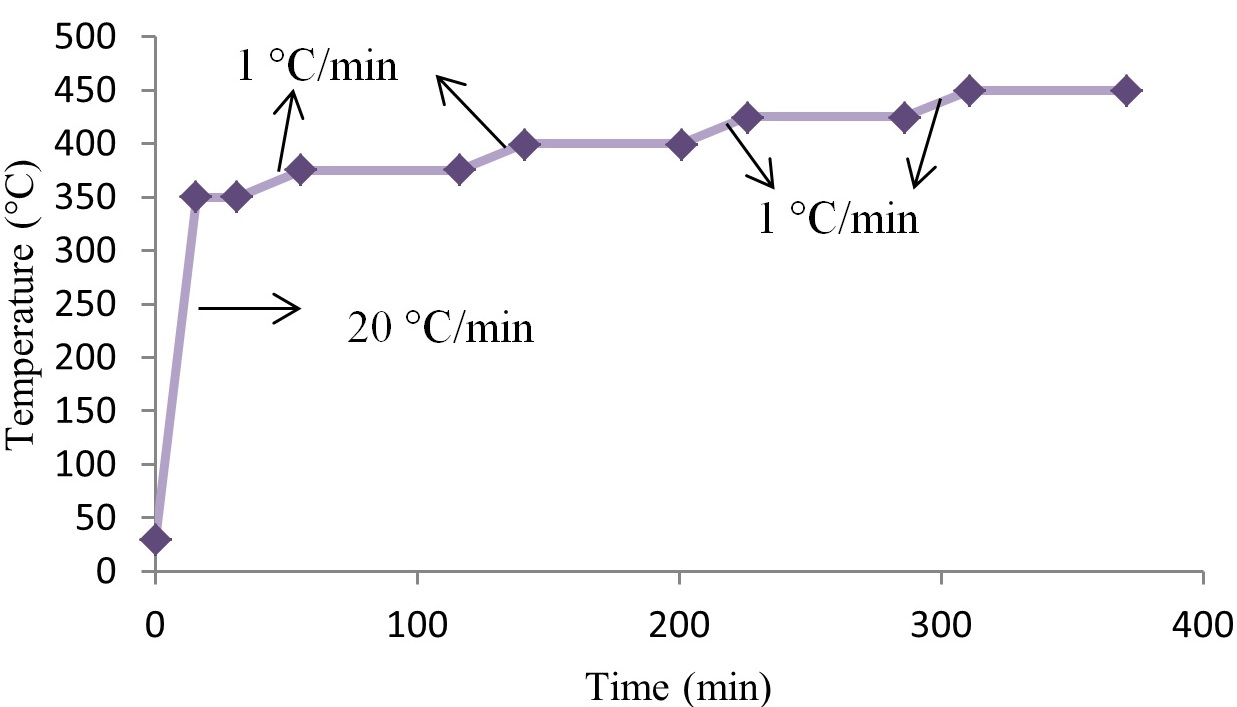}
     \caption[Polynorbornene]{The heating program for the decomposition
     of Unity 4698P} \label{heating} \end{figure}

    Removal of any polymer residue from the developed patterns was
    accomplished with a PlasmaTherm reactive ion etching (RIE) system
    using the following conditions: 45 sccm O2, 250 mTorr, 300 W at
    25\textdegree C. The etch rate of the polymer under these conditions
    was approximately 700 nm/min. Plasma enhanced chemical vapor
    deposition (PECVD) of the \ce{SiO2} overcoat was performed with a
    PlasmaTherm PECVD tool using the following conditions: 380 kHz RF,
    50 W power, 300\textdegree C, 550 mTorr, and a gas mixture of
    \ce{N2O} 1400 sccm and 2$\%$ \ce{SiH4} diluted in \ce{N2} 400 sccm.
    The \ce{SiO2} deposition rate was approximately 35 nm/min. The
    encapsulated sacrificial polymer structures were thermally
    decomposed in a Lindberg tube furnace purged with nitrogen.  It was
    found that a fast ramp rate, 20 \textdegree C/min, resulted in
    cavities with lower residue.  The heating cycle used for Unity 4698P
    deposition is shown in Fig. \ref{heating}.

\begin{figure}[htb] \centering \subfigure[Unity nanochannel]{
   \includegraphics[width=0.3\textwidth]{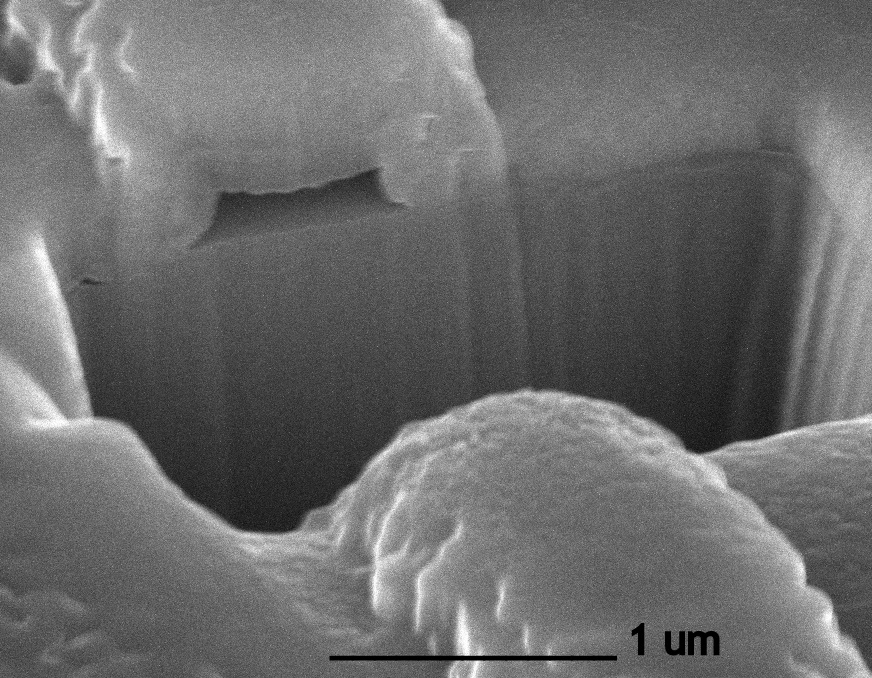} \label{unity1}
   }\\
   \subfigure[Sub 100nm thin channels with Unity]{
   \includegraphics[width=0.3\textwidth]{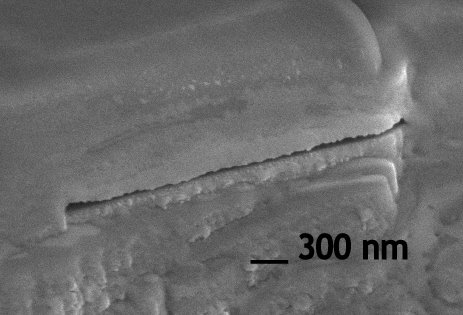} \label{unity2}
   }
   \caption[Channels fabricated with Unity]{Channels fabricated with
   Unity.  } \label{unity} \end{figure}

\begin{figure}[p] \centering \subfigure[High aspect ratio channel
   fabricated with Avatrel.]{
   \includegraphics[width=0.25\textwidth]{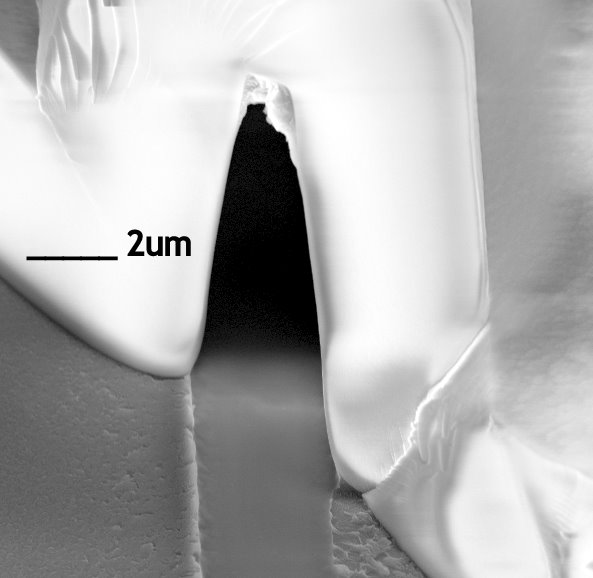} \label{av1}
   }\\
   \subfigure[Wider channel.]{
   \includegraphics[width=0.25\textwidth]{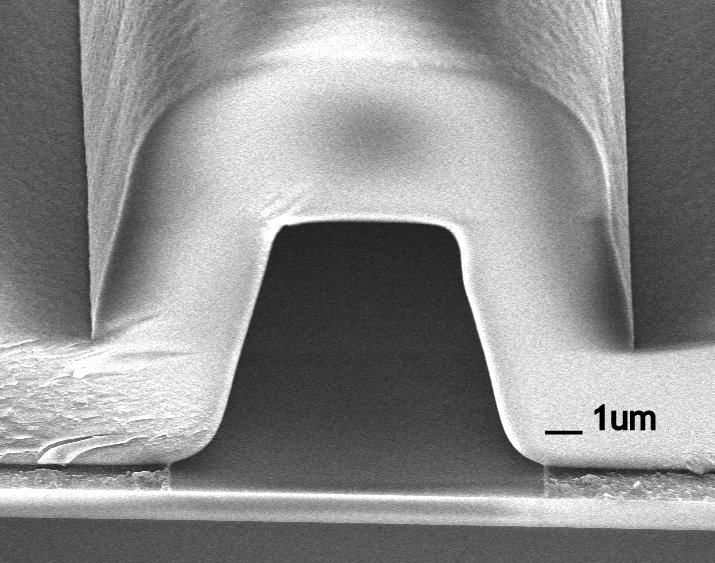} \label{av2}
   }\\
     \subfigure[If the thickness of the oxide cap is not large enough,
     thermal stress breaks the channels.]{
     \includegraphics[width=0.25\textwidth]{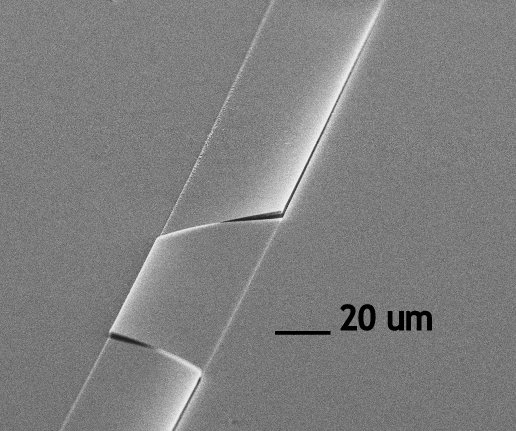}
     \label{av4}} \caption[Channels fabricated with Avatrel]{Channels
     fabricated with Avatrel using electron beam lithography.}
     \label{avatrel} \end{figure}

  One of the pertinent processing issues with \ce{SiO2} covered
  air-channels was found to be the temperature for deposition of the
  encapsulating layers. The oxide PECVD deposition temperature was
  limited to 300 \textdegree C. Above these temperatures, the overcoat
  material severely cracked. The most important factor limiting the
  deposition temperature of the films is the mismatch between the
  coefficient of thermal expansion of PNB (CTE = 127 ppm/\textdegree C)
  and \ce{SiO2} (CTE = 0.6 to 0.9 ppm/\textdegree C). Due to this
  mismatch, the overcoat films crack from stress developed between the
  substrate and the film during cooling from the deposition temperature
  \ref{av4}.   The final structure are shown in Fig. \ref{unity}

 The resulting structures can be seen in Fig. \ref{avatrel}. As it can
 be seen in this figure, there is no residue inside the channels, which
 is promising for optofluidic integration, as otherwise the performance
 of the optical devices would be highly degraded.

\section{Optofluidic integration} 
\begin{figure}[htb] \centering
\includegraphics[width=0.3\textwidth]{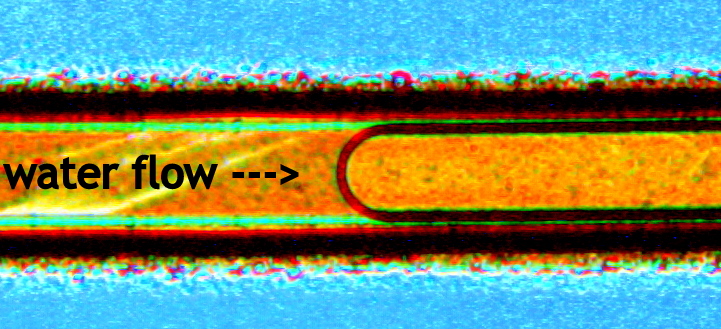} \caption[Flow of water
inside the channel.]{Flow of water inside the channel.} \label{flow1}
\end{figure}

After the channel fabrication is
optimized, integration with optical devices is the logical next step. We
needed to overcome several challenges for this process. The first
challenge was that the very first devices would not pass the fluid.
Further investigation showed that the decomposition process leaves the
inside surfaces of the channels hydrophobic. It is virtually impossible
to flow a liquid through a narrow hydrophobic channel. A high
temperature oxygen plasma in an asher makes  the channels hydrophilic
and the fluid flows inside the channels (as can be seen in Fig.
\ref{flow1}).

The next challenge was to make sure no residue is left inside the
channels (otherwise the performance of the optical devices would be
degraded). Usually for conventional photonic devices a piranha clean
removes the residues effectively. Piranha solution, also known as
piranha etch, is a mixture of sulfuric acid  and hydrogen peroxide, used
to clean organic residues off substrates. Because the mixture is a
strong oxidizer, it will remove most organic matter, and it will also
hydroxylate most surfaces (add OH groups), making them extremely
hydrophilic (water compatible). But because of the bubbles generated
during the piranha etch this process is not compatible with these
channels. Instead a solution of chromic acid is used. Chromic acid is a
mixture made by adding concentrated sulfuric acid to a dichromate, which
may contain a variety of compounds, including solid chromium trioxide.

\begin{figure}[htb] \centering
\includegraphics[width=0.3\textwidth]{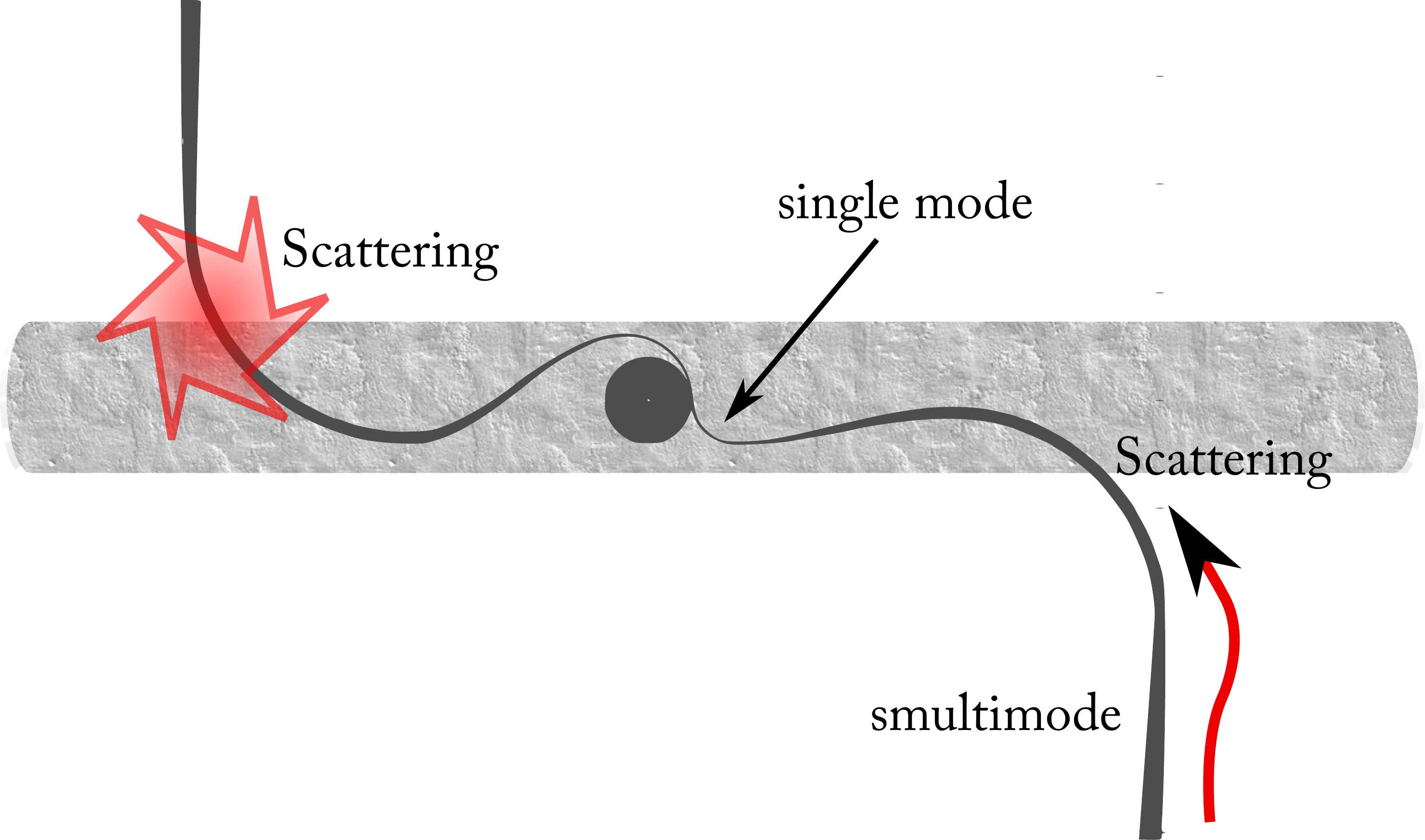} \caption[Disk and
channel integration.]{The size of the disk puts an upper limit on the
width of the channel. Scattering from the waveguide-channel interface
can be reduced by using a multimode waveguide at the interface.}
\label{scatter1} \end{figure}

Another issue to be considered is the scattering of light from the
waveguide-channel interface. As the index mismatch between the oxide
covered sections and the channel is large (specially if no fluid is
flowing inside the channel), a large portion of the optical guided wave
is scattered at the boundary (Fig. \ref{scatter1}). To alleviate this
problem we used large multi-mode   waveguides at the intersection and
tapered the waveguide to a single mode profile inside the channels.
Considering this issue, it is undesirable to cover a ring resonator with
the channels only partially, as this would lead to very low quality
factors due to the scattering.

As we need to cover the resonators with the channels, the width of the
channels should be larger than the diameter of the resonators used
(20-40 microns). This leads to fragile overcoats (Fig. \ref{av4}) unless
a thick oxide layer is used, which in turn leads to longer and higher
temperature decomposition conditions.

\section{Characterization} To characterize the fabricated structures,
the output light of a  tunable laser diode source (New
Focus\texttrademark   TLB-6305)    is coupled to the  cleaved facet of
the waveguide  using a Mitutoyo 20x long distance  objective lens. A
quarter wave plate and a polarizer ensure the light energy is in the TE
mode. The wavelength of the laser is swept across the 652--660 nm
wavelength range in 0.25 pm steps, and the   transmission is measured as
a function of wavelength by a Si detector at the waveguide  output. The
data is then transfered to the computer through a data acquisition (DAQ)
card. One long distant lens and one regular objective are used to
collect light from the top and the output respectively.

A custom-built microscope in the Z direction is used for most of the
analysis. Using 50/50 polarizing beam splitters, a spectrometer, a
detector and a camera are integrated in the microscope. For all the
setup structures Thorlabs 30 mm cage system where used. The cage system
is very versatile and stable and allowed fast and easy reconfigurations
in the setup without requiring an extensive alignment readjustment. The
only necessary part lacking to this day is a cage-mountable flip mirror
(which would increase the measured power by eliminating some of the beam
splitters.)

As a test of the capabilities of our system to measure fluorescence, the
output of a waveguide covered with 60 mg/lit Oxazine dye (Abs/Em at 
646/670 nm) is measured with an Ocean Optics spectrometer. A sharp edge
filter (with an optical density (OD) of over 70 is used to filter out
the pump). The unfiltered and filtered spectra are shown in Fig.
\ref{fl}.

\begin{figure}[hbp] \centering \subfigure[]{
\includegraphics[width=0.45\textwidth]{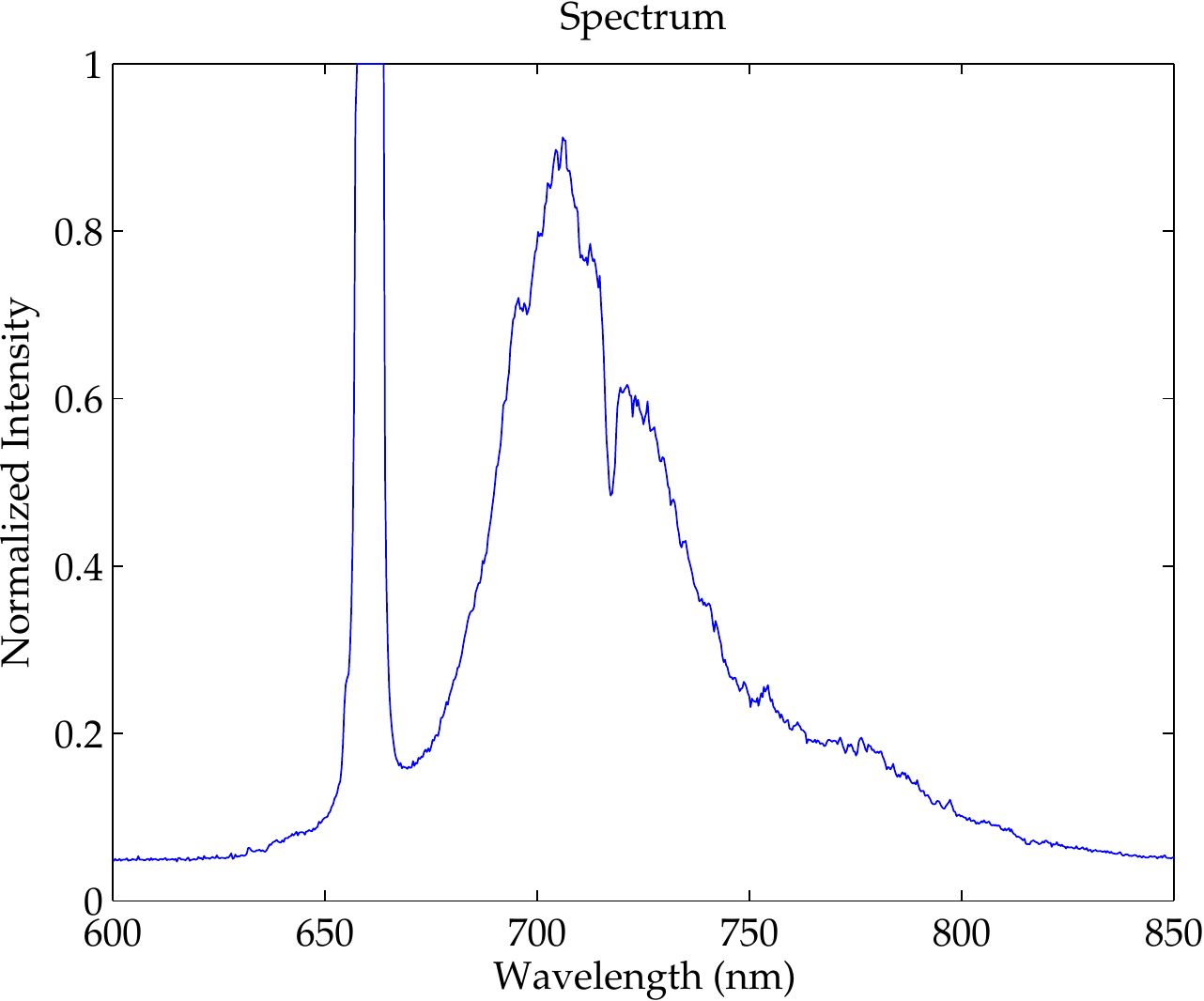} \label{fl1}
}\\
\subfigure[]{ \includegraphics[width=0.45\textwidth]{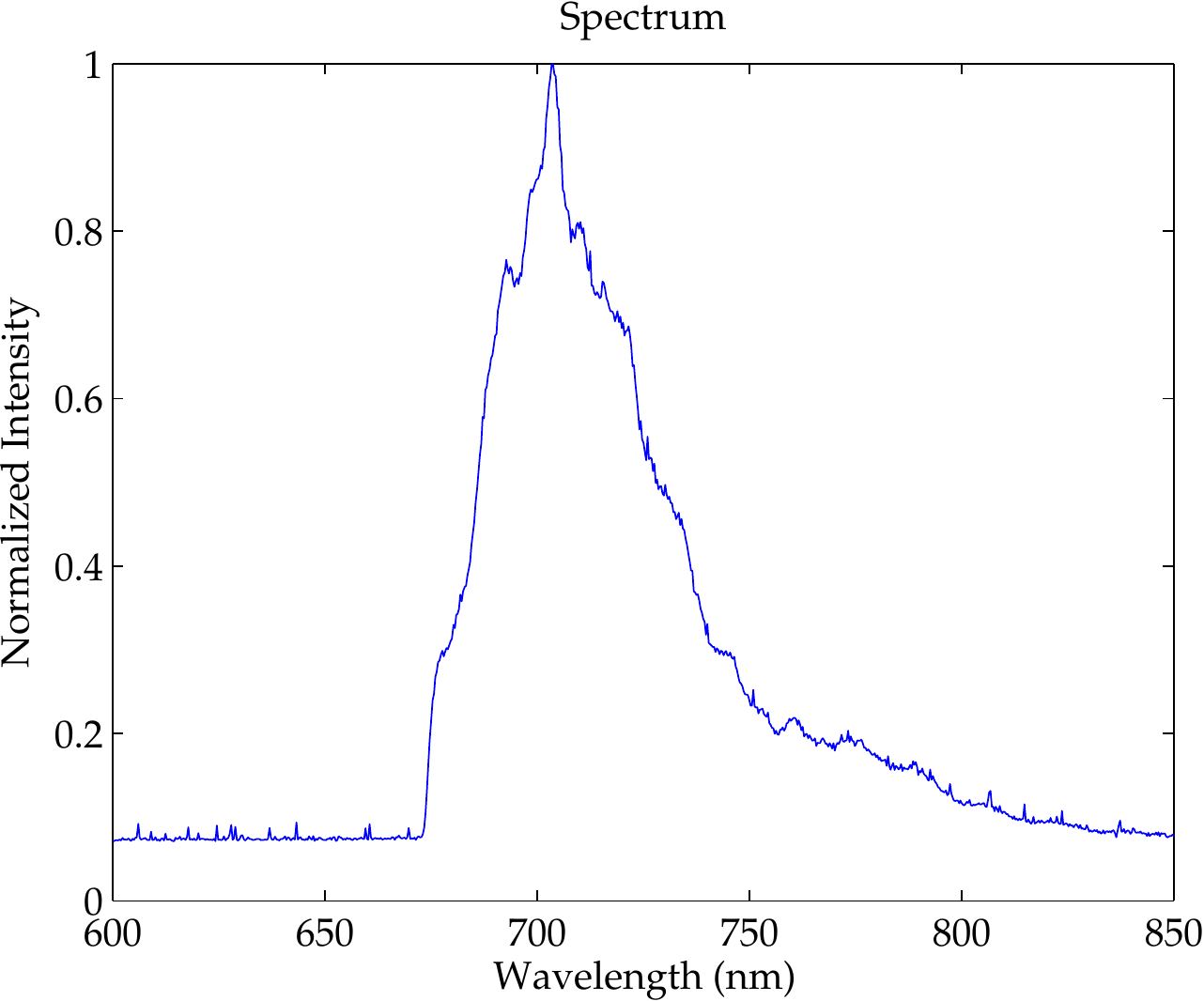}
\label{fl2}} \caption[Fluorescence coupled to a waveguide]{(a)
Fluorescence from Oxazine pumped and collected with SiN waveguides. (b)
The pump signal is filtered out.} \label{fl} \end{figure}

\begin{figure}[hbp] \centering \subfigure[]{
\includegraphics[width=0.45\textwidth]{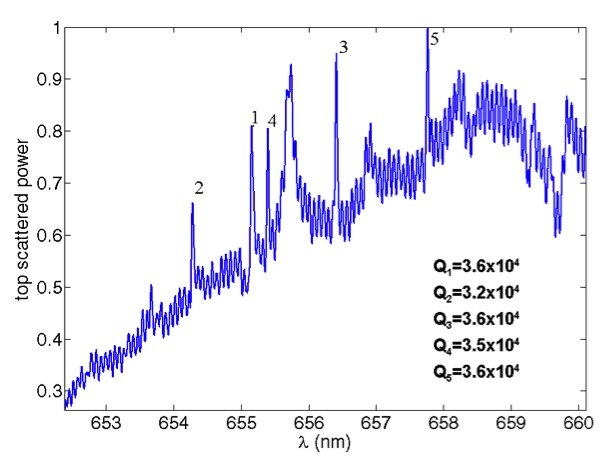} \label{fl1}
	}\\
	\subfigure[]{
	\includegraphics[width=0.45\textwidth]{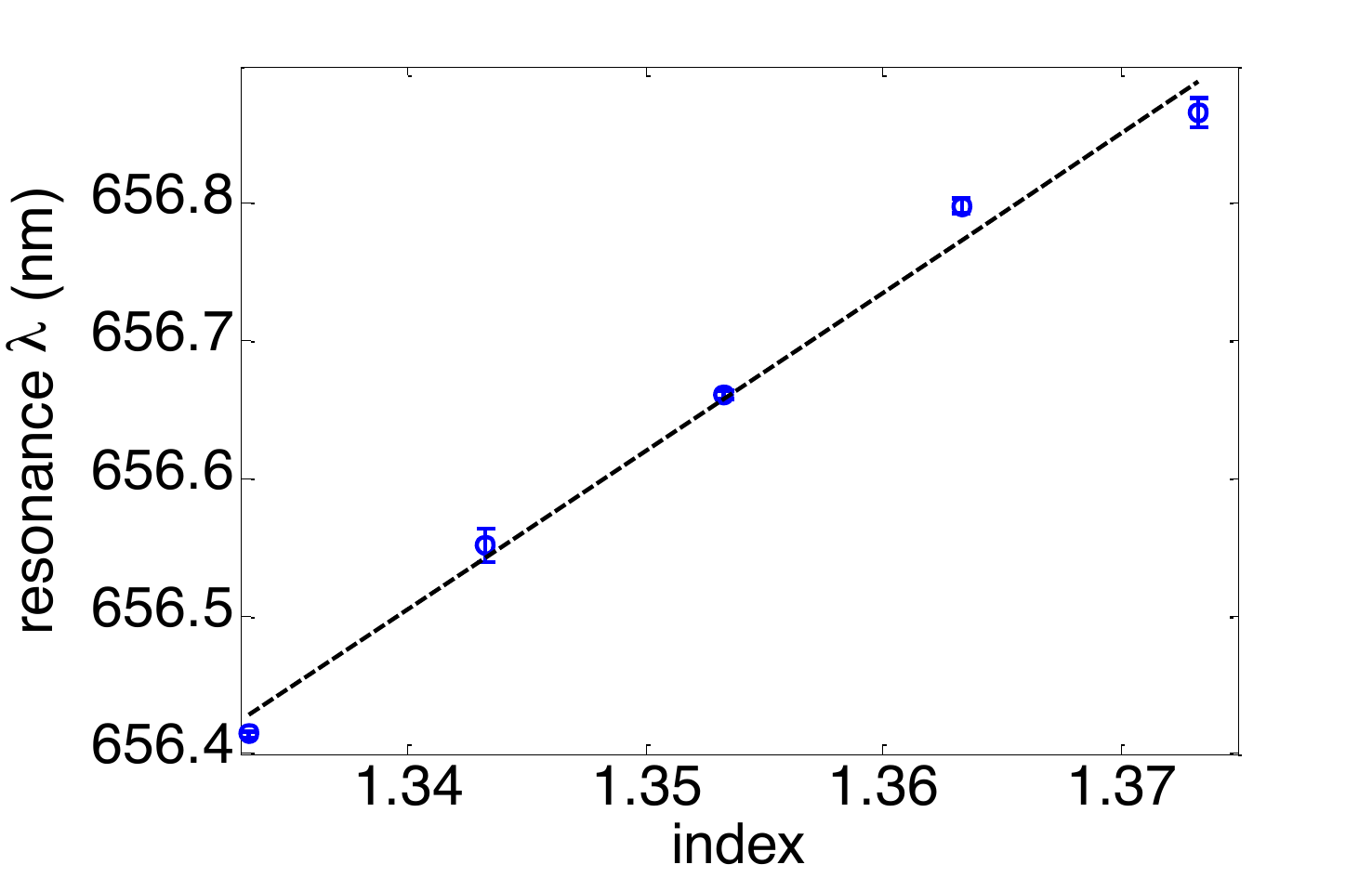}
	\label{fl2}} \caption[Resonance shift]{(a) High quality factor
	resonance modes of a silicon nitride cavity submerged in dextrose
	solution. (b)  Red-shift in resonance wavelength as the refractive
	index of the solution is increased by adding dextrose.} \label{fl}
	\end{figure}

\section{Conclusion} In this paper we showed that it is possible to
integrate micro/nanochannels with optical devices. The channels made
with low-temperature decomposability are the most promising for future
applications. Considering possible size reduction of the photonic
devices (specially through photonic crystal cavities discussed in this
thesis), it is possible to shrink the size of the channels even further
and achieve ultra-small sample sizes and multi-mode sensing
functionalities through florescence and Raman signals. The Raman signal
can be excited and collected through metallic nanoparticles fabricated
on top of the photonic devices and integrated with nanofluidic channels
for single molecule sensing.

\footnotesize{
\bibliography{document} 
\bibliographystyle{osajnl} 
}

\end{document}